\documentclass[twocolumn, resetfootnote, times]{aastex701}

\DeclareUnicodeCharacter{0301}{\'{}} 
\DeclareUnicodeCharacter{00E1}{\'{a}} 
\DeclareUnicodeCharacter{00E9}{\'{e}} 
\DeclareUnicodeCharacter{00ED}{\'{i}} 
\DeclareUnicodeCharacter{00F3}{\'{o}} 
\DeclareUnicodeCharacter{00FA}{\'{u}} 
\DeclareUnicodeCharacter{00E7}{\c{c}} 
\DeclareUnicodeCharacter{00E2}{\^{a}} 
\DeclareUnicodeCharacter{00EA}{\^{e}} 
\DeclareUnicodeCharacter{00F4}{\^{o}} 
\DeclareUnicodeCharacter{00E3}{\~{a}} 
\DeclareUnicodeCharacter{00F5}{\~{o}} 

\usepackage{multirow}
\usepackage{soul}

\begin{document}

\title{The rings of (2060) Chiron: Evidence of an evolving system}

\author[0000-0003-1000-8113]{C.~L.~Pereira}
\affiliation{Observat\'orio Nacional/MCTI, R. General Jos\'e Cristino 77, CEP 20921-400, RJ, Brazil}
\affiliation{Laborat\'orio Interinstitucional de e-Astronomia/LIneA, Av. Pastor Martin Luther King Jr, 126, Torre 3000 / sala 817. CEP 20765-000, RJ, Brazil}
\email{chrystianpereira@on.br}

\author[0000-0003-2311-2438]{F.~Braga-Ribas}
\affiliation{Federal University of Technology - Paran\'a (UTFPR-Curitiba), Rua Sete de Setembro, 3165, CEP 80230-901, PR, Brazil}
\affiliation{Laborat\'orio Interinstitucional de e-Astronomia/LIneA, Av. Pastor Martin Luther King Jr, 126, Torre 3000 / sala 817. CEP 20765-000, RJ, Brazil}
\affiliation{Observat\'orio Nacional/MCTI, R. General Jos\'e Cristino 77, CEP 20921-400, RJ, Brazil}
\email{felipebribas@gmail.com}

\author[0000-0003-1995-0842]{B.~Sicardy}
\affiliation{Laboratoire Temps Espace (LTE), Observatoire de Paris, Universit\'e PSL, CNRS UMR 8255, Sorbonne Universit\'e, LNE, 61 Av. de l'Observatoire, F75014 Paris, France}
\email{bruno.sicardy@obspm.fr}

\author[0000-0002-6477-1360]{R.~Leiva}
\affiliation{Instituto de Astrof\'isica de Andaluc\'ia, Glorieta de la Astronom\'ia S/N, 18008 Granada, Spain}
\email{rodleiva.astro@gmail.com}

\author[0000-0002-8211-0777]{M.~Assafin}
\affiliation{Universidade Federal do Rio de Janeiro - Observat\'orio do Valongo, Ladeira Pedro Ant\^onio 43, CEP 20080-090, RJ, Brazil}
\email{massaf@ov.ufrj.br}

\author[0000-0003-0088-1808]{B.~E.~Morgado}
\affiliation{Universidade Federal do Rio de Janeiro - Observat\'orio do Valongo, Ladeira Pedro Ant\^onio 43, CEP 20080-090, RJ, Brazil}
\email{bmorgado@ov.ufrj.br}

\author[0000-0002-8690-2413]{J.~L.~Ortiz}
\affiliation{Instituto de Astrof\'isica de Andaluc\'ia, Glorieta de la Astronom\'ia S/N, 18008 Granada, Spain}
\email{ortiz@iaa.es}

\author[0000-0002-1123-983X]{P.~Santos-Sanz}
\affiliation{Instituto de Astrof\'isica de Andaluc\'ia, Glorieta de la Astronom\'ia S/N, 18008 Granada, Spain}
\email{psantos@iaa.es}

\author[0000-0002-1642-4065]{J.~I.~B.~Camargo}
\affiliation{Observat\'orio Nacional/MCTI, R. General Jos\'e Cristino 77, CEP 20921-400, RJ, Brazil}
\affiliation{Laborat\'orio Interinstitucional de e-Astronomia/LIneA, Av. Pastor Martin Luther King Jr, 126, Torre 3000 / sala 817. CEP 20765-000, RJ, Brazil}
\email{camargo@on.br}

\author[0000-0002-2103-4408]{G.~Margoti}
\affiliation{Observat\'orio Nacional/MCTI, R. General Jos\'e Cristino 77, CEP 20921-400, RJ, Brazil}
\affiliation{Federal University of Technology - Paran\'a (UTFPR-Curitiba), Rua Sete de Setembro, 3165, CEP 80230-901, PR, Brazil}
\affiliation{Laborat\'orio Interinstitucional de e-Astronomia/LIneA, Av. Pastor Martin Luther King Jr, 126, Torre 3000 / sala 817. CEP 20765-000, RJ, Brazil}
\email{giulianomargoti@on.br}

\author[0000-0001-8641-0796]{Y.~Kilic}
\affiliation{Instituto de Astrof\'isica de Andaluc\'ia, Glorieta de la Astronom\'ia S/N, 18008 Granada, Spain}
\email{ykilic@iaa.es}

\author[0000-0002-4106-476X]{G.~Benedetti-Rossi}
\affiliation{Laborat\'orio Interinstitucional de e-Astronomia/LIneA, Av. Pastor Martin Luther King Jr, 126, Torre 3000 / sala 817. CEP 20765-000, RJ, Brazil}
\email{gugabrossi@gmail.com}

\author[0000-0003-1690-5704]{R.~Vieira-Martins}
\affiliation{Observat\'orio Nacional/MCTI, R. General Jos\'e Cristino 77, CEP 20921-400, RJ, Brazil}
\affiliation{Laborat\'orio Interinstitucional de e-Astronomia/LIneA, Av. Pastor Martin Luther King Jr, 126, Torre 3000 / sala 817. CEP 20765-000, RJ, Brazil}
\email{rvm@on.br}

\author[0000-0001-7126-4562]{T.~F.~L.~L.~Pinheiro}
\affiliation{Observat\'orio Nacional/MCTI, R. General Jos\'e Cristino 77, CEP 20921-400, RJ, Brazil}
\affiliation{UNESP - São Paulo State University, Grupo de Din\^amica Orbital e Planetologia, CEP 12516-410, SP, Brazil}
\email{franciscopinheiro@on.br}

\author[0000-0002-4939-013X]{R.~Sfair}
\affiliation{UNESP - São Paulo State University, Grupo de Din\^amica Orbital e Planetologia, CEP 12516-410, SP, Brazil}
\affiliation{LIRA, CNRS UMR8254, Observatoire de Paris, Universit\'e PSL, Sorbonne Universit\'e, Universit\'e Paris Cit\'e, CY Cergy Paris Universit\'e, 92190 Meudon, France}
\email{rafael.sfair@unesp.br}

\author[0000-0002-6085-3182]{F.~L.~Rommel}
\affiliation{Florida Space Institute (University of Central Florida), 12354 Research Parkway, Partnership 1, Orlando, FL 32826, USA}
\affiliation{Laborat\'orio Interinstitucional de e-Astronomia/LIneA, Av. Pastor Martin Luther King Jr, 126, Torre 3000 / sala 817. CEP 20765-000, RJ, Brazil}
\affiliation{Federal University of Technology - Paran\'a (UTFPR-Curitiba), Rua Sete de Setembro, 3165, CEP 80230-901, PR, Brazil}
\email{flavialuane.rommel@ucf.edu}

\author[0000-0002-3362-2127]{A.~R.~Gomes-Júnior}
\affiliation{Institute of Physics, Federal University of Uberl\^andia, Av. Jo\~ao Naves de \'Avila, CEP 38408-100, MG, Brazil}
\affiliation{Laborat\'orio Interinstitucional de e-Astronomia/LIneA, Av. Pastor Martin Luther King Jr, 126, Torre 3000 / sala 817. CEP 20765-000, RJ, Brazil}
\email{altairgomesjr@gmail.com}

\author[0000-0003-3452-1114]{R.~C.~Boufleur}
\affiliation{Laborat\'orio Interinstitucional de e-Astronomia/LIneA, Av. Pastor Martin Luther King Jr, 126, Torre 3000 / sala 817. CEP 20765-000, RJ, Brazil}
\email{rodrigo.boufleur@linea.org.br}

\author[0000-0001-5963-5850]{R.~Duffard}
\affiliation{Instituto de Astrof\'isica de Andaluc\'ia, Glorieta de la Astronom\'ia S/N, 18008 Granada, Spain}
\email{duffard@iaa.es}

\author[0000-0002-2193-8204]{J.~Desmars}
\affiliation{Institut Polytechnique des Sciences Avanc\'ees IPSA, 63 boulevard de Brandebourg, F-94200 Ivry-sur-Seine, France}
\affiliation{Laboratoire Temps Espace (LTE), Observatoire de Paris, Universit\'e PSL, CNRS UMR 8255, Sorbonne Universit\'e, LNE, 61 Av. de l'Observatoire, F75014 Paris, France}
\email{josselin.desmars@obspm.fr}

\author[0000-0003-4058-0815]{D.~Souami}
\affiliation{LIRA, CNRS UMR8254, Observatoire de Paris, Universit\'e PSL, Sorbonne Universit\'e, Universit\'e Paris Cit\'e, CY Cergy Paris Universit\'e, 92190 Meudon, France}
\affiliation{Department of Mathematics, naXys, University of Namur, Rue de Bruxelles 61, Namur 5000, Belgium}
\email{damya.souami@obspm.fr}

\author[0000-0003-0419-1599]{N.~Morales}
\affiliation{Instituto de Astrof\'isica de Andaluc\'ia, Glorieta de la Astronom\'ia S/N, 18008 Granada, Spain}
\email{nicolas@iaa.es}

\author{F.~Arrese}
\affiliation{Department of Physics, National University of La Pampa (UNLPAM), Santa Rosa, La Pampa, Argentina}
\email{arrese.fany@exactas.unlpam.edu.ar}

\author[0000-0003-1464-9276]{K.~Barkaoui}
\affiliation{Instituto de Astrof\'isica de Canarias (IAC), Calle V\'ia L\'actea s/n, 38200, La Laguna, Tenerife, Spain}
\affiliation{Astrobiology Research Unit, Universit\'e de Li\`ege, All\'ee du 6Ao\^ut 19C, B-4000 Li\`ege, Belgium}
\affiliation{Department of Earth, Atmospheric and Planetary Science, Massachusetts Institute of Technology, 77 Massachusetts Avenue, Cambridge, MA 02139, USA}
\email{khalid.barkaoui@iac.es}

\author[0000-0001-9892-2406]{A.~Burdanov}
\affiliation{Department of Earth, Atmospheric and Planetary Science, Massachusetts Institute of Technology, 77 Massachusetts Avenue, Cambridge, MA 02139, USA}
\email{burdanov@mit.edu}

\author{C.~A.~Colazo}
\affiliation{Observatorio Astron\'omico El Gato Gris (MPC I19) - Tanti, C\'ordoba, Argentina}
\affiliation{Observatorio de Ra\'ul Meli\'a Carlos Paz, Carlos Paz, C\'ordoba, Argentina}
\email{cacolazo@hotmail.com}

\author{C.~A.~Domingues}
\affiliation{Observat\'orio Estrela do Sul, Estrada Mar\'a, km 05, Gleba Ribeir\~ao Aquidaban, CEP 87112-390, PR, Brazil}
\email{cad96950@gmail.com}

\author{H.~Dutra}
\affiliation{Universidade Federal do Rio de Janeiro - Observat\'orio do Valongo, Ladeira Pedro Ant\^onio 43, CEP 20080-090, RJ, Brazil}
\email{hdutra@ov.ufrj.br}

\author{R.~C.~Gargalhone}
\affiliation{Laboratório Nacional de Astrof\'isica, Rua dos Estados Unidos, 154, CEP 37504-364, MG, Brazil}
\email{rgargalhone@lna.br}

\author{C.~Jacques}
\affiliation{SONEAR Observatory, Serra da Piedade, Caet\'e, MG, Brazil}
\affiliation{SONEAR2 Observatory, Rua Os\'orio de Morais 22, Bairro Ouro Preto, Belo Horizonte, MG, Brazil}
\affiliation{SONEAR3 Observatory, Serra da Piedade, Km3, Caet\'e, MG, Brazil}
\email{cjacqueslf@gmail.com}

\author{F.~Jablonski}
\affiliation{Instituto Nacional de Pesquisas Espaciais, Div. de Astrof\'isica. Av. dos Astronautas, 1758, CEP 12227-010, SP, Brazil}
\email{fjjablonski@gmail.com}

\author[0000-0003-3433-6269]{L.~Liberato}
\affiliation{Universit\'e C\^ote d'Azur, Observatoire de la C\^ote d'Azur, CNRS, Laboratoire Lagrange, Bd de l'Observatoire, CS 34229, 06304 Nice Cedex 4, France}
\email{luana.l.mendes.94@gmail.com}

\author{R.~Melia}
\affiliation{Observatorio de Ra\'ul Meli\'a Carlos Paz, Carlos Paz, C\'ordoba, Argentina}
\email{raulrobertomelia@gmail.com}

\author{J.~C.~Oliveira}
\affiliation{Independent researcher, Secretaria de Estado da Educaç\~ao do Paran\'a (SEED-PR), PR, Brazil}
\email{jc.oliveira64@gmail.com}

\author{M.~Sardiña}
\affiliation{Observatorio El Catalejo (MPC I48), Santa Rosa, La Pampa, Argentina}
\affiliation{Proyecto de Observación Colaborativa y Regional de Ocultaciones Asteroidales (POCROA), Argentina}
\affiliation{Grupo de Astronomía Pampeano (GAP), La Pampa, Argentina}
\email{marianwave@gmail.com}

\author{J.~Spagnotto}
\affiliation{Observatorio El Catalejo (MPC I48), Santa Rosa, La Pampa, Argentina}
\affiliation{Proyecto de Observación Colaborativa y Regional de Ocultaciones Asteroidales (POCROA), Argentina}
\affiliation{Grupo de Astronomía Pampeano (GAP), La Pampa, Argentina}
\email{jspagnotto@gmail.com}

\author[0009-0004-8376-5857]{T.~Speranza}
\affiliation{Observatorio Astron\'omico Municipal Reconquista (GORA OMR), Reconquista, Santa Fe, Argentina}
\email{donsperanza32@gmail.com}

\author{A.~Wilberger}
\affiliation{Observatorio Astron\'omico Los Cabezones (MPC X12), Santa Rosa, La Pampa, Argentina}
\affiliation{Grupo de Astronomía Pampeano (GAP), La Pampa, Argentina}
\email{ajwilberger@gmail.com}

\author[0009-0004-2504-7670]{M.~A.~Zorzan}
\affiliation{Edmond Halley Astronomy Club, Praça Ademar Bornia, PR, Brazil}
\email{maicozorzan14@gmail.com}

\author{L.~S.~Brito}
\affiliation{UNESP - São Paulo State University, Grupo de Din\^amica Orbital e Planetologia, CEP 12516-410, SP, Brazil}
\email{lucas.s.brito@unesp.br}

\author{J.~P.~Cavalcante}
\affiliation{UNESP - São Paulo State University, Grupo de Din\^amica Orbital e Planetologia, CEP 12516-410, SP, Brazil}
\email{juhpaiva0@gmail.com}

\author{T.~Q.~Costa}
\affiliation{Instituto Federal do Paran\'a, Campus Ivaipor\~a, Rua Max Arthur Greipel, 505, 86870-000, PR, Brazil}
\email{thiago.costa@ifpr.edu.br}

\author[0000-0001-5589-9015]{M.~Emilio}
\affiliation{Universidade Estadual de Ponta Grossa, Campus Uvaranas, Av. General Carlos Cavalcanti, 4748, CEP 84030-900, PR, Brazil}
\affiliation{Observat\'orio Nacional/MCTI, R. General Jos\'e Cristino 77, CEP 20921-400, RJ, Brazil}
\email{marcelo_emilio@yahoo.com}

\author{E.~Garcia-Migani}
\affiliation{Grupo de Ciencias Planetarias, Universidad Nacional de San Juan - CONICET, Argentina}
\email{egarciamigani@conicet.gov.ar}

\author[0000-0003-1462-7739]{M.~Gillon}
\affiliation{Astrobiology Research Unit, Universit\'e de Li\`ege, All\'ee du 6Ao\^ut 19C, B-4000 Li\`ege, Belgium}
\email{michael.gillon@uliege.be}

\author{E.~Gradovski}
\affiliation{Observat\'orio Nacional/MCTI, R. General Jos\'e Cristino 77, CEP 20921-400, RJ, Brazil}
\affiliation{Federal University of Technology - Paran\'a (UTFPR-Curitiba), Rua Sete de Setembro, 3165, CEP 80230-901, PR, Brazil}
\affiliation{Laborat\'orio Interinstitucional de e-Astronomia/LIneA, Av. Pastor Martin Luther King Jr, 126, Torre 3000 / sala 817. CEP 20765-000, RJ, Brazil}
\email{erosgradovski@ufpr.br}

\author[0000-0001-8923-488X]{E.~Jehin}
\affiliation{STAR Institute, Universit\'e de Li\`ege, All\'ee du 6 Ao\^ut 19C, B-4000 Li\`ege, Belgium}
\email{ejehin@uliege.be}

\author{V.~Lattari}
\affiliation{UNESP - São Paulo State University, Grupo de Din\^amica Orbital e Planetologia, CEP 12516-410, SP, Brazil}
\email{victor.lattari@unesp.br}

\author[0000-0002-0866-1802]{M.~Malacarne}
\affiliation{Federal University of Esp\'irito Santo, Av. Fernando Ferrari, 514, ES, Brazil}
\email{marcio.malacarne@gmail.com}

\author{L.~A.~Mammana}
\affiliation{El Leoncito Astronomical Complex (CASLEO), Av. Espa\~na 1512 Sur J5402DSP, San Juan, Argentina}
\affiliation{Faculty of Astronomical and Geophysical Sciences, National University of La Plata, Paseo del Bosque S/N. La Plata, B1900FWA, Argentina}
\email{lmammana@casleo.gov.ar}

\author{M.~Melita}
\affiliation{Instituto de Astronom\'ia y F\'isica del Espacio. (CONICET-UBA). Intendente Güiraldes S/N. CABA. C1428ZAA. Argentina}
\affiliation{Faculty of Astronomical and Geophysical Sciences, National University of La Plata, Paseo del Bosque S/N. La Plata, B1900FWA, Argentina}
\affiliation{Instiuto de Tecnologia e Ingeneria. UNAHUR. Vergara 2222, Hurlingham, Prov. de Buenos Aires, Argentina}
\email{melita@iafe.uba.ar}

\author[0009-0006-8255-6441]{W.~Melo}
\affiliation{Instituto de Aeron\'autica e Espaço: São Jos\'e dos Campos, SP, Brazil}
\email{wandeclayt@gmail.com}

\author{A.~J.~Ortiz}
\affiliation{Instituto Federal do Paran\'a, Campus Ivaipor\~a, Rua Max Arthur Greipel, 505, 86870-000, PR, Brazil}
\email{adriano.ortiz@ifpr.edu.br}

\author{P.~Quitral-Manosalva}
\affiliation{Departamento de F\'isica, Universidad Cat\'olica del Norte, Avenida Angamos 0610, Casilla 1280, Antofagasta, Chile}
\email{paola.quitral@ce.ucn.cl}

\author{G.~Ramon}
\affiliation{UNESP - São Paulo State University, Grupo de Din\^amica Orbital e Planetologia, CEP 12516-410, SP, Brazil}
\email{giovana.ramon@unesp.br}

\author{I.~Rodrigues}
\affiliation{UNIVAP, Universidade do Vale do Para\'iba, Av. Shishima Hifumi, 2911, S\~ao Jos\'e dos Campos, 12244-000, SP, Brazil}
\email{irapuan.rodrigues@gmail.com}

\author{L.~Vanzi}
\affiliation{Department of Electrical Engineering and Center of Astro Engineering, Pontificia Universidad Catolica de Chile, Av. Vicu\~na Mackenna 4860, 7820436 Macul, Regi\'on Metropolitana, Chile}
\email{lvanzi@ing.puc.cl}


\begin{abstract}

The centaur (2060)~Chiron has long been a candidate for hosting material in orbit, based on occultation, photometric, and spectroscopic data. Here, we present a multi-chord stellar occultation observed on 10~September~2023~UT that reveals new and complex structures surrounding Chiron. High-cadence light curves show multiple secondary events that are best explained (when compared with a multi-shell interpretation) with a system of three confined rings located at average radii of 273~km, 325~km, and 438~km, the outermost of which lies beyond Chiron’s Roche limit. The rings appear coplanar, with a mean pole orientation of $\lambda = 151^{\circ} \pm 4^{\circ}$ and $\beta = 20^{\circ} \pm 6^{\circ}$. A broader, disk-like structure extends from about 200 to 800~km, and a newly detected, faint feature is observed at $\sim$1,380~km. 
Chiron thus appears as the fourth small Solar System body known for hosting a ring system. Comparisons with previous occultation events that have occurred since 1994 show that these features are not permanent. With these observations, we may witness for the first time the ongoing formation and evolution of a ring system.

\end{abstract}



\section{Introduction} 

The discovery of rings around the Centaur (10199)~Chariklo~\citep{BragaRibas2014}, the dwarf planet (136108)~Haumea~\citep{Ortiz2017}, and the large trans-Neptunian object (50000)~Quaoar~\citep{Morgado2023,Pereira2023} has reshaped our understanding of how such structures can form and survive in the outer Solar System. These unexpected detections, often involving material located beyond the classical Roche limit, suggest a broader range of dynamical conditions under which rings can persist around small bodies. 
The diversity found in the rings around small bodies suggests that their formation must have been the result of a combination of factors, such as primordial collisions, catastrophic impacts, or ejection processes~\citep{Sicardy2025}, combined with a dynamically favorable environment around the central 
body~\citep{Salo2024,Sicardy2024,GiuliattiWinter2023}.

The Centaur (2060)~Chiron has long been known for photometric activity attributed to a surrounding coma~\citep{Hartmann1990,Bus1989,Meech1990,Duffard2002}. Stellar occultations since the 1990s have revealed transient narrow features~\citep{Bus1996,Elliot1995} and, more recently, ring-like structures consistent across multiple events in 2011, 2018, and 2022~\citep{Ruprecht2015,Sickafoose2020,Ortiz2015,Sickafoose2023,Ortiz2023}. The latter identified two dense features embedded in a broad envelope, with material extending to $\sim$580~km from Chiron's center.

In early 2021, Chiron experienced a sustained $\sim$1~mag brightening \citep{Dobson2021,Ortiz2023}, 
attributed to dust or ice ejection due to exposure of volatiles or ice phase transitions \citep{Dobson2024}. This is consistent with increased material inferred from the 2022 occultation. James Webb Space Telescope (JWST/NIRSpec) observations in July 2023 detected gaseous CH$_4$, CO$_2$, and solid CO, suggesting a complex environment comprising a nucleus, a coma, and a debris ring \citep{PinillaAlonso2024}. 

In this Letter, we report a multi-chord stellar occultation by Chiron on 10~September~2023 UT. High-cadence photometry reveals multiple narrow features consistent with a three-ring system, embedded in a broad disk-like structure and accompanied by a newly detected and yet to be confirmed external component.  The study of these structures sheds light on the processes governing their formation and long-term evolution. It also broadens our understanding of the dynamics of similar structures on other scales, such as planetary rings and circum-planetary disks.

\section{Observations and Data Analysis}

\begin{table}[ht]
    \centering
    \footnotesize
    \caption{Occulted star and Chiron's details.}
    \begin{tabular}{l c}
    \hline \hline
                                & Occulted star \\
    \hline
    Epoch                       &   2023-09-10 05:09:06.1 UTC \\
    Source ID                   &   GDR3 2578194139653336448    \\
    Star position               &  $\alpha_{\star}$ = 1$^{\mathrm{h}}$06$^{\mathrm{m}}$30$^{\mathrm{s}}$.376171 $\pm$ 0.124 mas     \\
\hspace{3.6mm} at Epoch$^1$     &  $\delta_{\star}$ = 8$^{\circ}$44'41''.750877 $\pm$ 0.126 mas   \\
    Magnitudes$^2$              &  G: 13.26, B: 13.87, V: 13.23, R: 13.08,      \\
                                &  J: 12.03, H: 11.65, K: 11.610 \\
    Star apparent               &  0.0208 mas / 0.27 km    \\
\hspace{4.3mm} diameter$^3$         & \\
\hline
                                &   (2060) Chiron \\
    \hline
Ephemeris                       &   NIMAv15 \\
Geocentric distance             &   17.891995 au\\
Rotation period$^4$             &  5.917813 $\pm$ 0.000007 h\\
Equivalent radius$^5$           &  98 $\pm$ 17 km\\
Mass$^5$                        &  (4.8 $\pm$ 2.3) $\times10^{18}$~kg \\
\hline
    \end{tabular}
\begin{minipage}{0.45\textwidth}
\footnotesize
\textbf{Notes.} $^1$The star position was taken from the Gaia Data Release 3 (GDR3) star catalog \citep{GaiaColaboration2023} and is propagated to the event epoch using SORA \citep{GomesJr2022}. $^2$J, H, and K from the NOMAD catalog \citep{Zacharias2004}. $^3$Estimated using the empirical relations of \citet{Kervella2004}; the value of 0.27~km corresponds to its projected size at Chiron’s geocentric distance.\\
\textbf{References.} $^4$\citet{Marcialis1993}. $^5$\citet{BragaRibas2023}.
\end{minipage}
    \label{tab:star_chiron_information}
\end{table}

This work presents a joint analysis of four stellar occultations by Chiron between 2011 and 2023. Predictions and observational details for the 2011, 2018, and 2022 events are available in previous studies \citep[see;][]{Ruprecht2015, Sickafoose2020, Sickafoose2023, Ortiz2023}. The most recent event, on 10 September 2023, was predicted by the \textit{Lucky Star} project\footnote{\url{https://lesia.obspm.fr/lucky-star}} using Gaia Data Release 3 \citep{GaiaColaboration2023} stellar position and a high-precision \textsc{nima}\footnote{Numerical Integration of the Motion of an Asteroid} \citep{Desmars2015} ephemeris for Chiron (see Table \ref{tab:star_chiron_information} for details). A total of 31 observation sites across South America participated, coordinated via the Occultation Portal\footnote{\url{https://occultationportal.org}} \citep{Kilic2022}. The most significant data were obtained from the Perkin-Elmer 1.6-m telescope (hereafter PE160) at the Pico dos Dias Observatory (LNA, Brazil), with high-cadence (10 Hz) imaging. 

These observations revealed both a broad feature and embedded dense, confined structures. 
Additional secondary flux drops were detected by other stations in this event, aiding the determination of the ring pole orientation and semi-major axis, though they were unresolved. 
The 2023 occultation led to six effective detections of Chiron's solid body. While a full analysis of the main body’s shape and the fitted limb ellipse has been performed, this will be presented in a separate study. In the present work, we focus solely on the detection of secondary features and the center of the projected ellipse, which provides the offset used to model the ring system.

\section{Extended and Confined Ring Structures}
\begin{figure*}[t]
    \centering
    \includegraphics[width=0.98\linewidth]{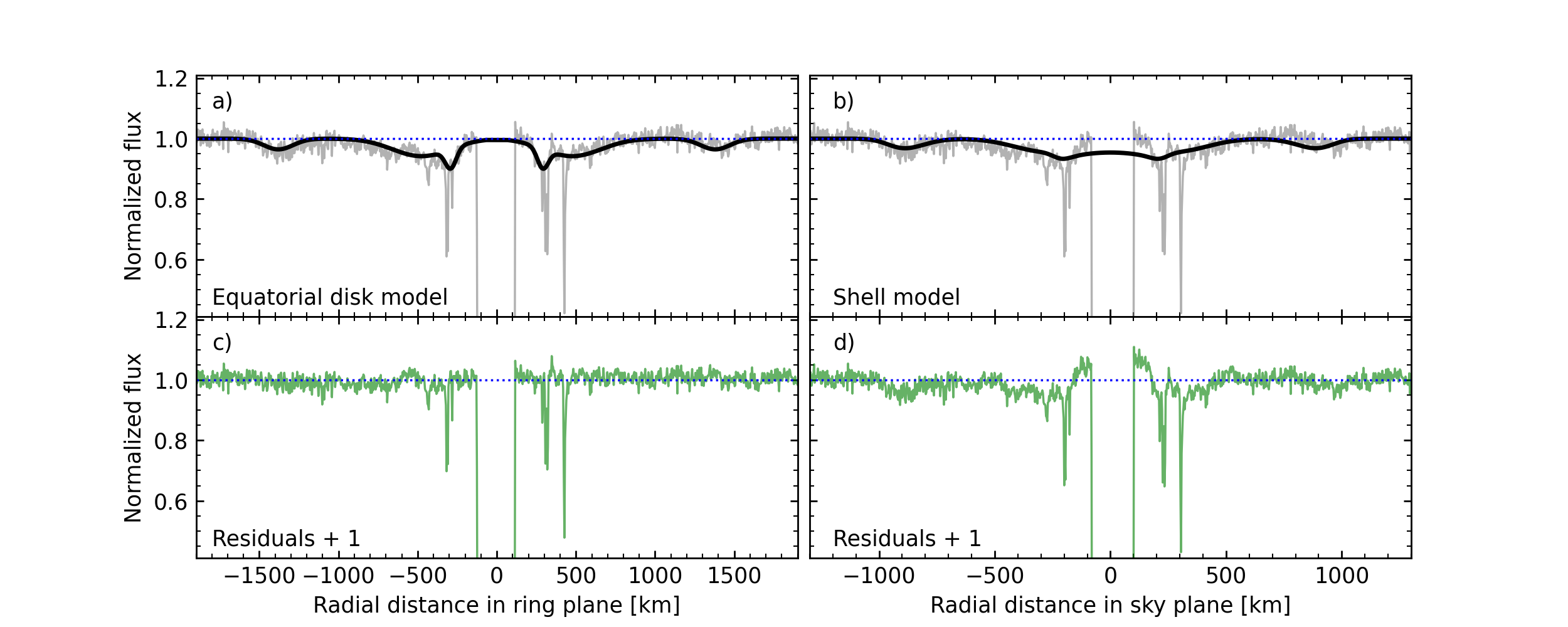}
    \caption{Comparison between the equatorial disk and shell models. In the panels, the black curve represents the correspondent model compared to the observed light curve. a) The time-based light curve is projected in the radial distance in the ring plane considering the preferred mean pole orientation. b) The normalized flux as a function as the radial distance in the sky plane, compared with the model based in the consecutive spherical shells. The residuals plus 1 for the equatorial disk (c) and shell (d) models}.
    \label{fig:2023_PE160_ring_gaussian}
\end{figure*}
The presence of a coma around Chiron and broad flux drops observed in past occultations motivated an investigation into whether extended material could be detected in the light curves. The 2023 occultation revealed sharp drops of flux embedded in broader, diffuse variations, along with a more distant structure around 1,380 km from Chiron’s center. 
To test the possible geometries of the extended material, we considered two simplified hypotheses: (i) a broad equatorial disk sharing the rings’ pole orientation, and (ii) multiple overlapping spherical shells. The latter were not intended as a physical model of Chiron’s dust coma, which is known to exist, but rather as a parametric representation of isotropic, diffuse extinction that could mimic the appearance of cometary comae. 
This approach allowed us to assess whether an isotropic distribution of dust could explain the data, without introducing the complexity of a full cometary coma model.

For both models, the broad flux drops were described by Gaussian profiles characterized by their central position $x_0$, width $\sigma$, and apparent opacity $p'$. The free parameters were optimized by minimizing residuals with a Levenberg–Marquardt algorithm, excluding the sharp flux drops of the confined rings by restricting the fit to data within $\pm2\sigma$ of the median baseline.

The equatorial disk model assumes a flat, circular structure with azimuthally uniform optical depth in Chiron’s equatorial plane, with the stellar flux mapped into radial distance profiles on the ring plane. In contrast, the spherical shell model was constructed from overlapping three-dimensional Gaussian functions projected onto the sky plane, intended as a parametric test of isotropic extinction rather than a physical model of the coma. The Gaussian models and the residuals are shown in Figure~\ref{fig:2023_PE160_ring_gaussian}, highlighting that only the equatorial disk provides a satisfactory fit to the data.

\begin{figure*}[t]
    \centering
    \includegraphics[width=0.475\linewidth]{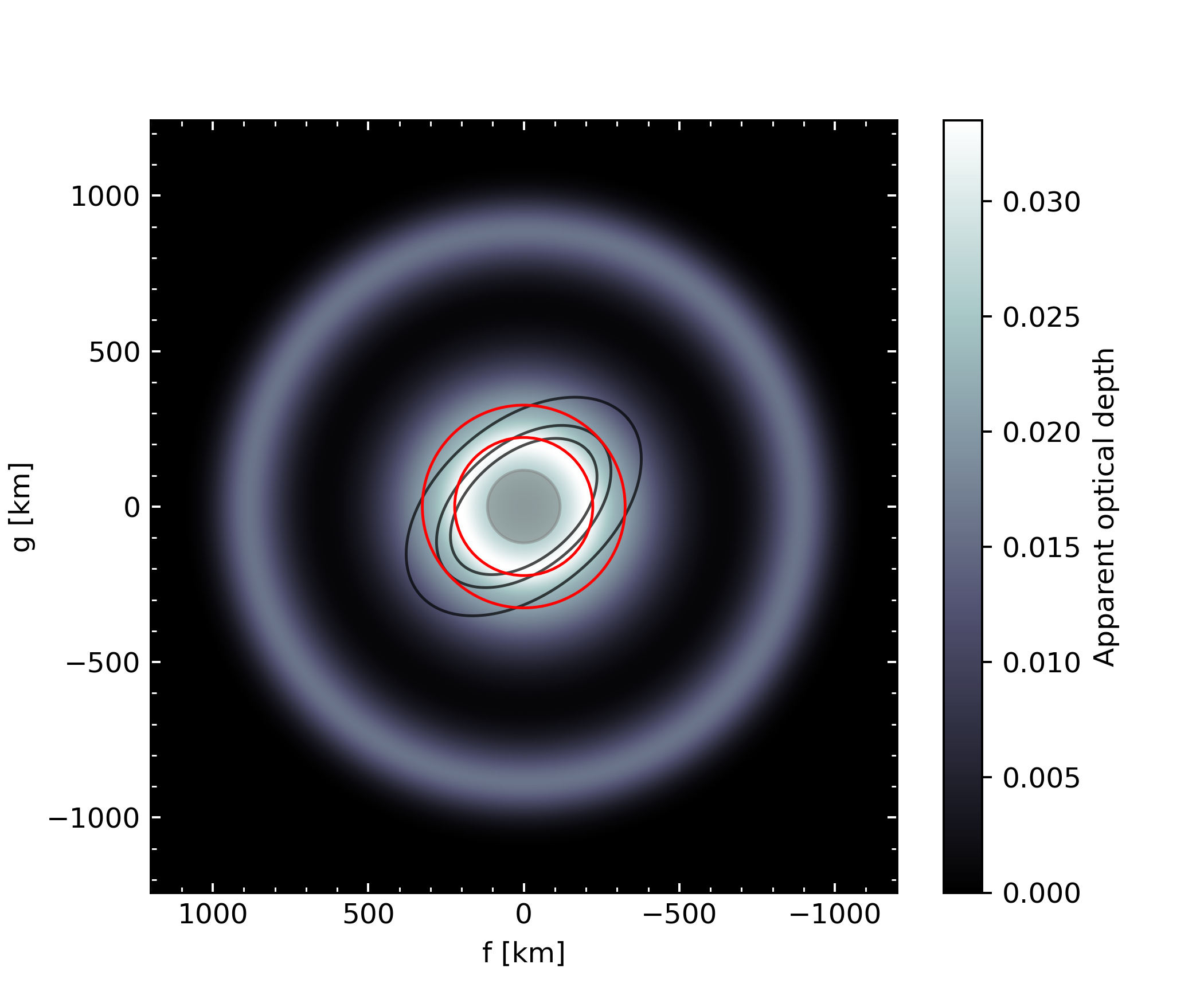}
    \includegraphics[width=0.48\linewidth]{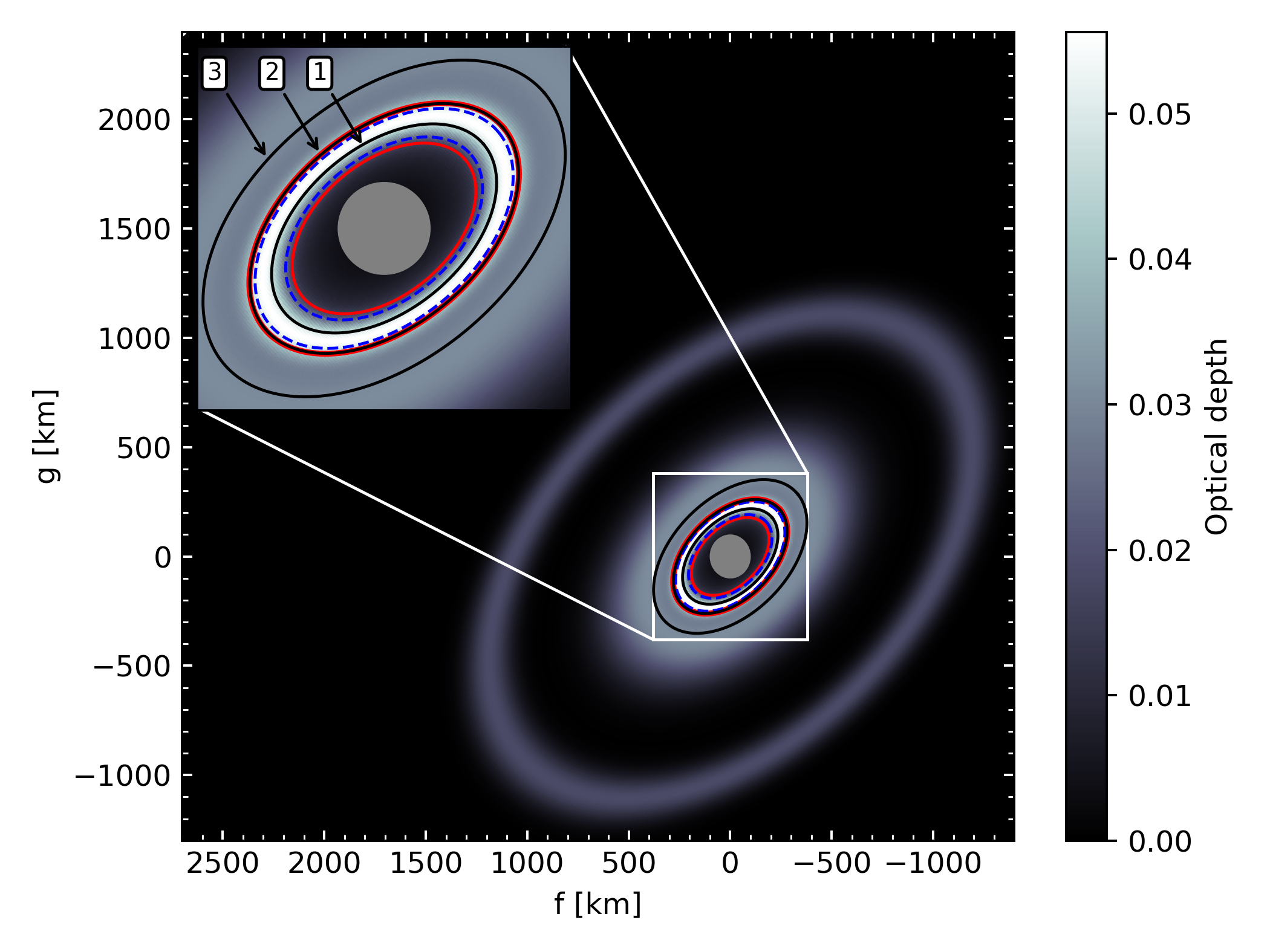}

    \caption{\textit{Left:} Sky-plane projection of the shell model. The color map shows the optical depth $\tau = - \ln(T)/2$ from three Gaussian-distributed spherical shells. The central grey circle (98~km radius) marks Chiron's projected size; black ellipses indicate the confined rings; red circles denote the Roche limits. \textit{Right:} Sky-plane view of the equatorial disk model. Optical depth $\tau = \tau_{\mathrm{N}}/|\sin(B)|$ is shown. The grey circle (98~km radius) marks Chiron. Black ellipses show the three narrow rings; dashed blue ellipses mark the 1/2 and 1/3 spin-orbit resonances; red ellipses indicate Roche limits. Inset: zoom-in of the innermost region with rings Chi1R (1), Chi2R (2), and Chi3R (3).}

    \label{fig:gaussian_shell}
\end{figure*}

We modeled Chiron's ring structures with routines based on SORA \citep{GomesJr2022} using square-well profiles, which provide the ingress and egress times and opacities of semi-transparent features. To isolate the confined rings, we first subtracted the extinction caused by the broad disk and fit the remaining structures in the residual light curves using $\chi^2$ minimization (Figure \ref{fig:2023_PE160_ring_gaussian}, panel c). Some features required composite models with multiple square boxes to achieve satisfactory fits (see Figure \ref{fig:gaussian_shell}). The radial distances of each structure were derived from the midpoints of ingress and egress times projected onto the ring plane, while radial widths were obtained from their separation. Optical properties, including apparent and normal opacities and optical depths, were computed from square-well fits accounting for Fresnel diffraction by the entire ring and Airy diffraction by individual particles \citep{Cuzzi1985,Roques1987}. 

We model Chiron’s rings as flattened circular structures centered on the body. For each secondary feature detected in the 2011, 2018, 2022, and 2023 occultations, we calculate its sky-plane position by averaging the disappearance and reappearance times. Associated uncertainties consider timing errors and the duration of each occultation. A projected ellipse is then fitted to these positions, taking into account the opening and position angles of the rings at each event. Assuming a stable pole orientation between 2011 and 2023, we searched for the best-fitting solutions for each ring’s radius and pole orientation, identifying three distinct rings. Each ring yields two mirrored solutions, with the preferred one consistent with observed long-term brightness variations. 

\section{Results}

The 2023 event reveals a broad disk-like feature with about 550~km in radial extent, with a median normal optical depth $\tau_{\mathrm{N}} = 0.016$, embedding at least three distinct dense rings: Chi1R, Chi2R, and Chi3R. The observed gradual return of the stellar flux near the central body to its baseline is consistent with a broad disk, rather than with a shell configuration (Figure~\ref{fig:gaussian_shell}). 

\begin{figure*}[t]
    \centering
    \includegraphics[width=0.75\linewidth]{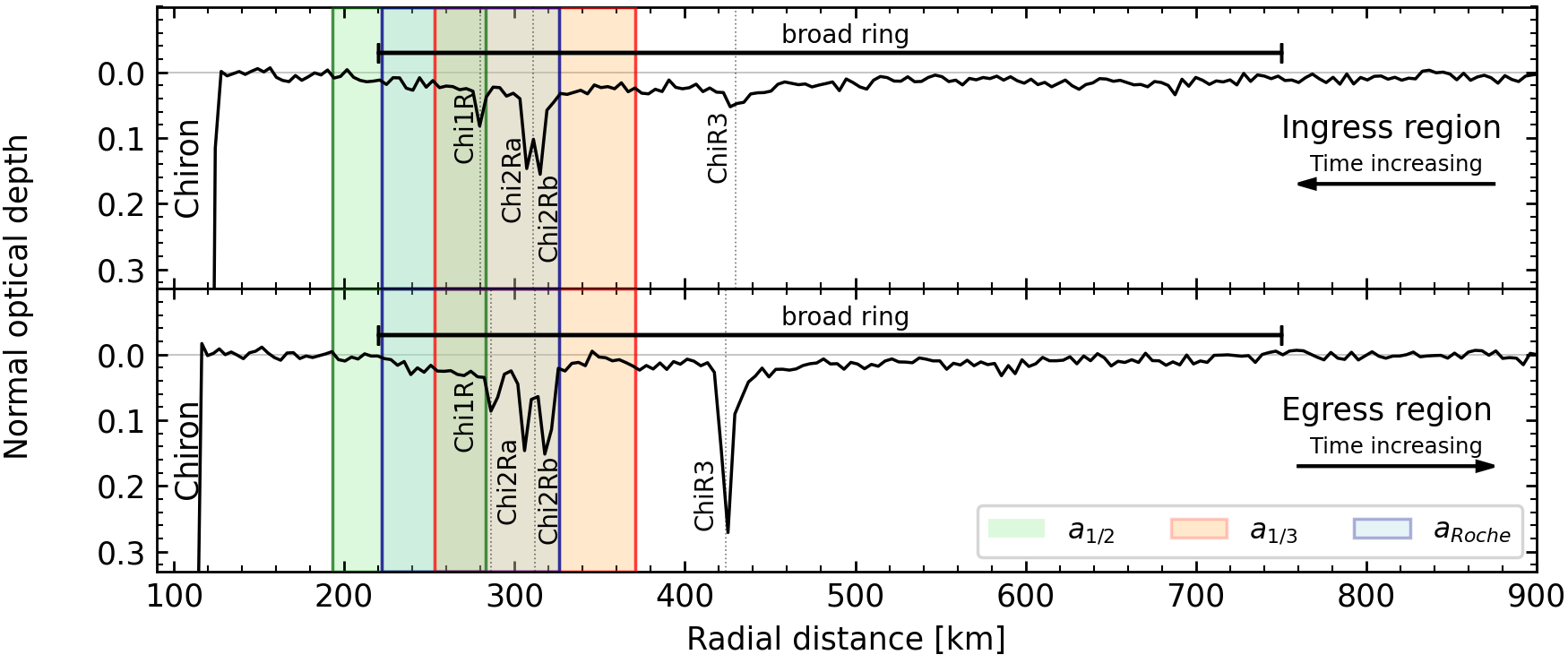}
    \caption{Occultation profile from the PE160 observation on September 10, 2023 event of the normal optical depth as a function of radial distance in the ring plane shows the extent of the broad ring and the arrangement of the dense rings. The Coloured regions indicate the boundaries of the SOR 1/2 resonance (green), SOR 1/3 resonance (red), and the Roche limit (blue). The vertical dotted lines indicates the respective ring center.}
    \label{fig:2023_tau_vs_radial_loc}
\end{figure*}
The innermost ring, Chi1R, is detected at $273 \pm 14$~km from Chiron's center. Marginally seen in 2011 \citep{Ruprecht2015,Sickafoose2020}, it now exhibits $\tau_{\mathrm{N}} = 0.045$–$0.12$ and equivalent width $E_p = 0.2$–$0.8$~km. Chi2R, located at $325 \pm 11$~km, consists of two components separated by a gap of 2–9~km (varying with longitude), with $\tau_{\mathrm{N}}$ ranging from 0.1 (2023) to 0.35 (2011) \citep{BragaRibas2023}. Its equivalent width varies from 0.6 to 0.9~km across events.
A third ring, Chi3R, lies at $\sim$438~km, and presents characteristics reminiscent of Quaoar’s Q1R ring \citep{Morgado2023, Pereira2023}, Saturn’s F ring \citep{Murray2018}, and Neptune’s arcs \citep{Hubbard1986}. Its optical depth varies azimuthally: from $\tau_{\mathrm{N}} = 0.03$ (tenuous, ingress) to 0.3 (dense, egress). The radial width spans from 7 to 44~km. Assuming the dense portion detected at egress is a confined arc, we estimate a minimum azimuthal extent of $\sim$25$^\circ$, or $\sim$190~km. Combining with 2022 detections, the inferred extension may reach up to 140$^\circ$, corresponding to an arc of $\sim$1,100~km.
An outermost symmetrical feature, Chi4R, is detected at $1,380 \pm 11$~km. It is modeled as a broad Gaussian profile with radial width of $93 \pm 11$~km and peak $\tau_{\mathrm{N}} \sim 10^{-3}$.

\begin{figure*}[t]
    \centering
    \includegraphics[width=1\linewidth]{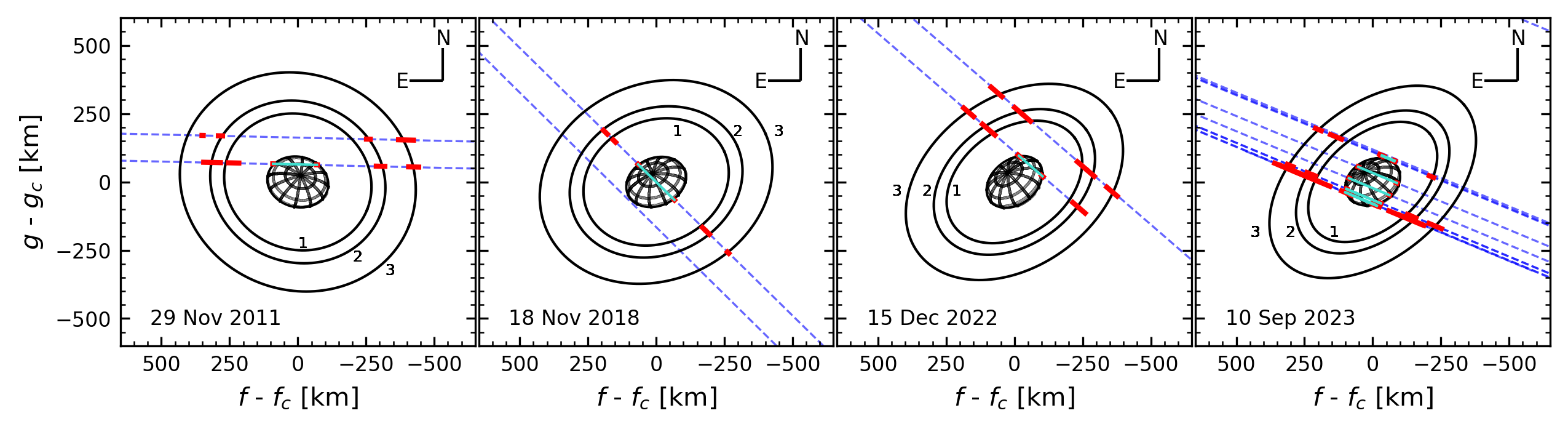}
    \caption{Sky-plane projection of Chiron and its confined rings. The projected ellipsoid and the occulting chords for each event is plotted in turquoise, with their $1\sigma$ uncertainties represented by the red segments at the chord extremities. The dashed blue lines represent the occulting chords as seen from each site. The red segments mark the projected widths of the ring events plus their respective $1\sigma$ error bars. The labels 1, 2, and 3 refers to the Chi1R, Chi2R, and Chi3R rings, respectively.}
    \label{fig:Final_Chiron_mean_pole}
\end{figure*}

Fitting all secondary features across events yields three co-planar, circular rings centered on Chiron (Figure \ref{fig:Final_Chiron_mean_pole}). Each ring’s pole is consistent within $1\sigma$ uncertainties with a mean orientation of $\lambda = 151.3^\circ \pm 4.0^\circ$, $\beta = 19.9^\circ \pm 6.0^\circ$, in ecliptic coordinates. This solution is preferred because it better accounts for the variations in Chiron's absolute magnitude \citep{Ortiz2015}. The alternative, mirrored solution is $\lambda = 331.5^\circ \pm 4.2^\circ$, $\beta = -20.1^\circ \pm 5.9^\circ$.

Recent estimates suggest that Chiron has a density of $\rho_{\rm C} = 1{,}119 \pm 4$~kg,m$^{-3}$, corresponding to a mass of $\mathrm{M}_{\rm C} = (4.8 \pm 2.3) \times 10^{18}$~kg \citep{BragaRibas2023}. Based on this mass, we estimate that the synchronous orbit is located at $a_{\mathrm{sync}} = 150 \pm 29$~km using a rotation period of $\mathrm{P} = 5.917813 \pm 0.000007$~h \citep{Marcialis1993}. 
To within uncertainties, the Chi1R and Chi2R rings align with the 1/2 and 1/3 spin-orbit resonances (SORs), located at $a_{1/2} = 238 \pm 45$~km and $a_{1/3} = 312 \pm 59$~km, respectively (Figure~\ref{fig:2023_tau_vs_radial_loc}). At these radii, orbiting particles complete one revolution around Chiron for every two and three rotations of the body, respectively. 
The Roche critical density indicates the minimum density a body must have to stay gravitationally bound against the tidal forces of the central body. Classically, if the material's density is lower than this critical value, it tends to disperse and form a ring, whereas if higher, it may coalesce into a stable satellite \citep[See;][]{Morgado2023}.
Assuming $\gamma = 1.6$ for tidally deformed particles, the Roche critical density $\rho_{\mathrm{Roche}}$ is 413~kg\,m$^{-3}$, 245~kg\,m$^{-3}$, and 104~kg\,m$^{-3}$ for Chi1R, Chi2R, and Chi3R, respectively. If the ring material density is $\sim$400~kg\,m$^{-3}$ (comparable to Saturnian moons), the classical Roche limit ranges from 274–338~km depending on the ring particle's shape, spin state, and cohesion \citep{Sicardy2025}. The dense Chi1R and Chi2R rings may thus lie within the Roche limit, while the more distant Chi3R and Chi4R fall outside it.

\section{Discussion} \label{sec:discussion}

The multi-year consistency of detections of confined material strongly argues against interpreting the structures as transient phenomena, such as cometary jets or ephemeral dust shells. On the other hand, observational data indicate that the broad disk was not detected around Chiron in 2011 and 2018. Its inclusion in photometric models \citep[e.g.][]{Ortiz2015} would yield absolute magnitudes significantly brighter than observed, unless its albedo is anomalously low. This discrepancy suggests that the broad disk formed recently, likely within the last decade.
The region with the highest concentration of material (within the broad disk) spans from $258$~km to $328$~km, matching the locations of the SOR 1/2 and SOR 1/3. Models on Chariklo's stability and ring confinement suggest that the region inside the SOR 1/2 would be cleared on short timescales (less than 10 years) due to the elongation of the central body \citep{GiuliattiWinter2023,Sicardy2019}. Given Chiron’s triaxial shape \citep{BragaRibas2023}, similar clearing is expected. This suggests that any material that was dispersed around Chiron through some mechanism would be expected to follow the same dynamical evolution.

We can estimate the broad ring mass from its surface density $\Sigma = (4/3) \tau_{\mathrm{N}} r \rho_{\rm part}$, where $r = 10^{-2}~\mathrm{m}$ is the assumed particle radius, $\rho_{\rm part} = 400\,\mathrm{kg\,m}^{-3}$ is bulk density of the particles, and $\tau_{\mathrm{N}} = 0.016$ is the median of the normal optical depth in the ring region. We adopt $r = 10^{-2}~\mathrm{m}$ as this value lies at the lower end of the size range observed in Saturn’s main rings \citep{Cuzzi2018} and provides a conservative estimate in the absence of constraints for Chiron. For a ring with radius $a_{\mathrm{ring}} = 485$~km and radial width $\mathrm{W}_{r} \approx 600$~km, we found $m_{\mathrm{ring}} \sim 10^{11}$~kg and a ring-to-Chiron mass-ratio of $\mu \approx 3.5 \times 10^{-8}$.  If released at a cometary mass-loss rate of 5,000~kg~s$^{-1}$, as estimated for comet 29P/Schwassmann-Wachmann-1 \citep{Jewitt2009}, this mass could be produced in under one year, comparable to the timescale of Chiron’s 2021 outburst \citep{Dobson2024}. 
It is important to note that the comet 29P orbits much closer to the Sun than Chiron, which was near aphelion during the 2021 outburst event. This implies that the same level of mass loss at Chiron's distance would require more efficient energy deposition or a different activity mechanism. 

The formation of an equatorial dust disk as a by-product from outburst was observed in the comet C/2014 B1 (Schwartz), which exhibits a stable, disk-shaped coma \citep{Jewitt2019}. 
The morphology of the structure is consistent with dust grain ejection with a speed similar to the escape velocity, assisted by the centripetal forces that favor equatorial ejections \citep{Jewitt2019}.  A similar mechanism may operate on Chiron, particularly as a new hemisphere has been progressively illuminated since 2000. This exposure could trigger localized sublimation, leading to episodic dust ejection.

Alternatively, Chiron’s activity may result from impacts with debris generated by the fragmentation of another object. This hypothesis is supported by the temporal coincidence between Chiron’s activity episodes and its passages through descending nodes \citep{Ortiz2023}. The object 2015~RD$_{277}$ has been proposed as a potential progenitor of such a debris swarm \citep{GilHutton2024}, though the statistical likelihood of this scenario remains unconstrained.

Another plausible source of dispersed material is the gradual fragmentation of a small satellite orbiting Chiron. Long-term deviations in Chiron’s absolute magnitude since 2010 suggest the presence of additional reflective material in its environment \citep{Duffard2002, Dobson2024}. Supporting this interpretation, JWST/NIRSpec observations in 2023 detected gas-phase CH$_4$ and CO$_2$, but no CO \citep{PinillaAlonso2024}. Given CO’s lower sublimation temperature, its absence indicates that activity is not driven by continuous solar heating of surface or near-surface layers. Instead, CO likely remains sequestered in amorphous ices or buried deeper in the subsurface, inaccessible to insolation. This points to episodic exposure of volatile-rich material, such as might occur through satellite disruption, as a more viable driver of recent activity.

Assuming that all the dispersed material originated from the catastrophic breakup of a single spherical satellite with a density of $\rho_{\rm part} = 400$~kg~m$^{-3}$, the equivalent radius of that progenitor body would be approximately 0.5~km. Alternatively, adopting the bulk density of Chiron ($\rho_{\rm Chiron} = 1{,}119$~kg~m$^{-3}$ \citep{BragaRibas2023}), the disrupted satellite would have a radius of approximately $0.3$~km. However, attributing Chiron's recent brightening solely to the disruption of a satellite poses a challenge when considering its long history of recurrent outbursts over the past 70 years \citep{Bus2001, Duffard2002,Ortiz2015,Dobson2024}. This discrepancy suggests that either multiple fragmentation events have occurred, or that additional mechanisms such as internal pressure build-up of subsurface volatiles may be periodically driving activity on Chiron. Thus, we may be observing the result of different sources of dust material feeding Chiron's vicinity.

In the context of recent ring formation around Chiron, the presence of material resulting from a sudden event can lead to particles rapidly settling into the equatorial plane of the central body. This process is driven by gravitational perturbations caused by Chiron's dynamical oblateness $J_2$, the influence of radiation pressure, and mutual collisions between particles \citep{Goldreich1982,Marzari2020}. These collisions dissipate energy, promoting the alignment of particle orbits and contributing to the formation of a flattened equatorial ring structure. The collisional damping process can occur over short timescales, within fewer than ten orbital periods of a particle located near the central body’s Roche limit \citep{Ida1997,Kokubo2000}. 

An important implication of our findings is that we may be witnessing the ongoing formation of a ring system around a small body. The current features observed around Chiron, including the presence of dust, gas-phase volatiles, and confined ring material, could represent a transitional stage of a circum-object environment. This stage may bridge the gap between an initially chaotic distribution of debris (possibly generated by satellite fragmentation, transient outgassing events, or collisional processes) and the eventual development of a narrow, dense, and highly flattened ring system. Chiron may thus represent a rare observational window into an intermediate evolutionary phase, offering a potential missing link in the formation pathway of ring systems around small Solar System bodies.
\begin{acknowledgments}
We thank the anonymous referee for helpful comments that improved the manuscript.
CLP thanks the FAPERJ/DSC-10 E-26/204.141/2022, FAPERJ/PDR-10 E-26/200.107/2025, and FAPERJ 200.108/2025. 
This work was carried out within the Lucky Star umbrella that agglomerates the efforts of the Paris, Granada, and Rio teams, which is funded by the European Research Council under the European Community H2020 (ERC Grant Agreement No. 669416). 
This study was financed in part by the Coordenação de Aperfeiçoamento de Pessoal de Nível Superior – Brasil (CAPES) – Finance Code 001 and the National Institute of Science and Technology of the e-Universe project (INCT do e-Universo, CNPq grant 465376/2014-2). 
The following authors acknowledge the respective grants: 
FB-R: CNPq 316604/2023-2; 
BS acknowledges support by the French ANR project Roche, number ANR-23-CE49-0012;
MA: CNPq 427700/2018-3, 310683/2017-3, and 473002/2013-2, and FAPERJ E-26/210.705/2024; 
JIBC: CNPq 305917/2019-6 and 306691/2022-1, and FAPERJ 201.681/2019;
RV-M: CNPq 307368/2021-1; 
GM: CAPES PhD scholarship 88887.015375/2024-00;
FLR: CNPq 103096/2023-0, Florida Space Institute’s Space Research Initiative, and the University of Central Florida’s Preeminent Postdoctoral Program (P3); 
ARG-Jr: FAPEMIG APQ-02987-24;
IR: CNPq 314267/2021-2;
GR: FAPESP 2024/20150-1;
RL acknowledges the support of Instituto de Estudios Astrofísicos, Universidad Diego Portales, for providing telescopes and logistical assistance;
Part of this work was supported by the Spanish projects PID2020-112789GB-I00 from AEI and Proyecto de Excelencia de la Junta de Andalucía PY20-01309; Authors JLO, PS-S, NM, and RD acknowledge financial support from the Severo Ochoa grant CEX2021-001131-S funded by MCIN/AEI/ 10.13039/501100011033;
PS-S and YK acknowledge financial support from the Spanish I+D+i project PID2022-139555NB-I00 (TNO-JWST);
Funding for KB was provided by the European Union (ERC AdG SUBSTELLAR, GA 101054354).
We thank observers L.~A.~Pereira, O.~M.~Ribas, O.~Margoti, R.~Gil-Hutton, J.~Rodrigues, E.~F.~Morato, A.~J.~L.~Pereira, C.~Andreolla, N.~Castro, M.~Fidêncio~Neto, E.~Miglioranza, F.~Ruzza, A.~Stechina, E.~Townsend for their efforts in collecting data.
Some results were based on observations taken at the 1.6 m telescope on Pico dos Dias Observatory of the National Laboratory of Astrophysics (LNA/Brazil). 
Based on observations collected at the European Organisation for Astronomical Research in the Southern Hemisphere under ESO programme 111.24SV.002.
This research made use of SORA, a Python package for stellar occultations reduction and analysis, developed with the support of ERC Lucky Star and LIneA/Brazil. 
This work has made use of data from the European Space Agency (ESA) mission Gaia (https://www.cosmos.esa.int/gaia), processed by the Gaia Data Processing and Analysis Consortium (D.P.A.C., https://www.cosmos.esa.int/web/gaia/dpac/consortium). 
TRAPPIST is funded by the University of Liège and the Belgian Fund for Scientific Research (Fond National de la Recherche Scientifique, FNRS) under the grant PDR T.0120.21. E.J. G is F.R.S.-FNRS Research Director.
The ULiege's contribution to SPECULOOS has received funding from the European Research Council under the European Union’s Seventh Framework Programme (FP/2007-2013; grant Agreement no. 336480/SPECULOOS), from the Balzan Prize and Francqui
Foundations, from the Belgian Scientific Research Foundation (F.R.S.-FNRS; grant no. T.0109.20), from the University of Liège, and from the ARC grant for Concerted Research Actions financed by the Wallonia-Brussels Federation.
We thank Chalé e Camping Luz (Sapopema, Paraná, Brazil) for kindly providing the site for our observation.
\end{acknowledgments}

\begin{contribution}
C.L.P. wrote the paper and designed the figures, with major contributions from F.B.-R., B.S., and B.E.M.
F.B.-R., R.L., J.L.O., P.S.-S., and R.D. organized the stellar occultation observational campaigns, with the collaboration of Y.K. and N.M.
C.L.P. analyzed the data and derived the physical parameters presented in the paper, with inputs from M.A., G.M., B.E.M., F.B.-R., and J.I.B.C.
C.L.P., A.R.G.-J., R.C.B., G.B.-R., F.L.R., G.M., M.A., Y.K., and J.D. were the main developers of the data analysis software used in this project.
F.B.-R., M.A., B.E.M., P.S.-S., J.I.B.C., R.V.-M., T.F.L.L.P., and D.S. contributed to the interpretation and review of the results.
All remaining authors participated in the observation of the occultation event. They are listed in alphabetical order—first those who reported positive and negative results, followed by those who experienced overcast conditions or technical failures.
All authors had the opportunity to review the results and comment on the manuscript.

\end{contribution}

%
\facility{Observatório do Pico dos Dias (OPD/LNA)}

\software{SORA (\url{https://github.com/riogroup/SORA}); PRAIA (\url{https://ov.ufrj.br/en/PRAIA/}), \textsc{astropy} \citep{Astropy2013,Astropy2018,Astropy2022}}
\bibliography{my}{}
\bibliographystyle{aasjournalv7}

\appendix
\section{Observation circumstances.} The observational circumstances for each site involved in the stellar
occultation campaign presented in this work are listed in Table \ref{tab:obscircums_20230910}.
\begin{table}[ht]
        \footnotesize
        \centering
        \caption{Observation circumstances for all the observatories involved in the September 10, 2023 event.}
        \label{tab:obscircums_20230910}
        \vspace{4mm}
            \begin{tabular}{c c c c c} \hline \hline
                                                &   \textbf{Latitude ($^\circ$ ' '')}    &   \textbf{Telescope (mm)}  &   \textbf{Exposure (s)}    &               \\ 
                    \textbf{Site}                        &   \textbf{Longitude ($^\circ$ ' '')}   &   \textbf{Camera}          &   \textbf{Cycle (s)}       &   \textbf{Observers}   \\ 
                                                &   \textbf{Altitude (m)}                &   \textbf{Filter}          &   \textbf{Status}          &               \\ 
            \hline  
            OPD PE      & -22 32 05.886         & 1,600      & 0.120  &          \\ 
            Minas Gerais& -45 34 59.222         & IXon      & 0.133  & G. Benedetti-Rossi \\ 
            Brazil      & 1,864.0               & Clear     & Positive     &          \\   
            \hline
            
            OPD B\&C/IAG& -22 32 04.428         & 600       & 0.670     &   L. Liberato       \\ 
            Minas Gerais& -45 34 57.692         & IXon      & 0.683     & H. Dutra      \\ 
            Brazil      & 1,864.0               & Clear    & Positive  &                \\ 
            \hline

            OPD T40     & -22 32 06.910         & 400       & 1.000     &  F. Jablonski        \\ 
            Minas Gerais& -45 34 59.691         & Andor Zyla 4.2     & 1.110     & R. C. Gargalhone      \\ 
            Brazil      & 1,864.0               & Clear    & Positive  &                \\ 
            \hline
            
            OAMR        & -29 09 17.809         & 254                   & 0.800     &          \\ 
            Reconquista & -59 38 45.762         & Player One Ceres-M    & 2.000     & T. Speranza \\ 
            Argentina   & 50.0                  & Clear                 & Positive &          \\  
            \hline
            
                        & -29 08 25.183         & 305       & 0.600     &          \\ 
            Reconquista & -59 38 36.609         & QHY174GPS & 0.600     & A. Stechina \\ 
            Argentina   & 50.0                  & Clear     & Positive  &          \\ 
            \hline
            
            OES         & -23 30 49.164         & 200           & 0.500     & F. Braga-Ribas          \\ 
            Sarandi     & -51 51 33.270         & Raptor        & 0.500     & C. A. Domingues \\ 
            Brazil      & 469.0                 & Clear         & Positive     &          \\ 
            \hline

            OES - eVscope 1      & -23 30 49.164         & 114           & 1.000     & F. Braga-Ribas          \\ 
            Sarandi     & -51 51 33.270         & Sony IMX224LQR& 1.000     & M. Zorzan \\ 
            Brazil      & 469.0                 & Clear         & Positive     &          \\ 
            \hline
            
            Mobile      & -30 55 08.300         & 200           & 0.750     & J. Spagnotto          \\ 
            La Rioja    & -66 07 55.500         & QHY174M GPS   & 0.750     & M. Sardiña \\ 
            Argentina   & 455.0                 & Clear         & Positive  &          \\ 
            \hline

            Mobile      & -31 34 15.426         & 200           & 0.750     & A. Wilberger          \\ 
            La Rioja    & -66 14 11.256         & QHY174M GPS   & 0.750     & F. Arrese  \\ 
            Argentina   & 405.0                 & Clear         & Positive     &          \\ 
            \hline
            SONEAR3             & -19 49 27.263 & 450           & 2.000     &           \\ 
            Minas Gerais        & -43 41 24.025 & QHY600        & 4.240     & C. Jacques \\ 
            Brazil              & 1,830.0       & Clear         & Positive     &          \\ 
            \hline

            Villa Carlos Paz    & -31 24 48.499 & 254           & 1.000     & R. Meliá  \\ 
            Córdoba             & -64 30 21.499 & QHY174M GPS   & 1.000     & C. A. Colazo \\ 
            Argentina           & 708.4         & Clear         & Positive  &          \\ 
            \hline

            SONEAR2             & -19 52 55.082 & 280           & 3.000     &           \\ 
            Minas Gerais        & -43 49 03.120 & QHY600        & 5.740     & C. Jacques \\ 
            Brazil              & 885.0         & Clear         & Negative     &          \\ 
            \hline

            \multicolumn{5}{l}{\textbf{OPD:} Observatório do Pico dos Dias; \textbf{PE:} Perkin-Elmer; \textbf{B\&C:} Boller \& Chivens; \textbf{OAMR:} Observatorio}\\
            \multicolumn{5}{l}{Astronómico Municipal de Reconquista; \textbf{OES:} Observatório Estrela do Sul; \textbf{SONEAR:} Southern Obser-}\\
            \multicolumn{5}{l}{vatory for Near Earth Asteroids Research.}
      
\end{tabular}

            \end{table}

\setcounter{table}{1}
    \begin{table}[ht]
        \footnotesize
        \caption{[Cont.] Observation circumstances for all the observatories involved in the September 10, 2023 event.}
        \centering
        \vspace{4mm}
                \begin{tabular}{c c c c c} \hline \hline
                                                &   \textbf{Latitude ($^\circ$ ' '')}    &  \textbf{ Telescope (mm)}  &   \textbf{Exposure (s)}    &               \\ 
                    \textbf{Site}                        &   \textbf{Longitude ($^\circ$ ' '')}   &   \textbf{Camera}          &  \textbf{ Cycle (s) }      &   \textbf{Observers}   \\ 
                                                &  \textbf{ Altitude (m)}                &   \textbf{Filter }         &   \textbf{Status }         &               \\ 
                    \hline
            VLT UT4             & -24 37 37.301 & 8,400         & 0.100     &           \\ 
            Cerro Paranal       & -70 24 14.062 & HAWKI         &  N/A     & R. Leiva \\ 
            Chile               & 2,637.0       & J             & Negative     &          \\ 
            \hline
            SPECULOOS-S         & -24 36 57.917 & 1,000         & 0.500         &             \\ 
            Cerro Paranal       & -70 23 26.016 & Andor iKon-L  &  N/A          & A. Burdanov  \\ 
            Chile               & 2,479.2    & SDSS-$g'$ and $r'$ & Negative    & K. Barkaoui   \\ \hline
            
            OUC                 & -33 16 08.936 & 500           & 1.000     & N. Castro        \\ 
            Santiago            & -70 32 04.135 & Raptor Merlin & 1.000     & L. Vanzi   \\ 
            Chile               & 1,475.2       & Clear         & Overcast  & R. Leiva    \\ \hline
            
            CASLEO              & -31 47 55.018 & 2,153         & 2.000     & R. Gil-Hutton    \\ 
            San Juan            & -69 17 44.236 & CCD Directo   &  N/A      & E. Garcia-Migani  \\ 
            Argentina           & 2,485.0       & Clear         & Overcast  & L. Mammana        \\ \hline
            
            NTT                 & -29 15 32.054 & 3,580         & 0.100     & ULTRACAM team    \\ 
            La Silla            & -70 44 01.637 & ULTRACAM      & N/A         & R. Leiva    \\ 
            Chile               & 2,345.4       & $u'$, $g'$, $r'$  & Overcast  &     \\ \hline

            Alianza S4          & -31 47 12.401 & 450           & N/A         & J. L. Ortiz, N. Morales    \\ 
            San Juan            & -69 18 24.480 & SBIG STX11000 &  N/A        & P. Santos-Sanz   \\ 
            Argentina           & 2,664.0       & Clear         & Overcast  & R. Duffard   \\ \hline

            Dois Vizinhos       & -25 42 16.592 & 203           &  N/A        & F.Rommel, F.Ruzza  \\ 
            Paraná              & -53 05 51.360 & Raptor Merlin &  N/A        & E. Miglioranza  \\ 
            \multirow{2}{*}{Brazil}              & \multirow{2}{*}{550.0}         & \multirow{2}{*}{Clear}         & Overcast  & C. Andreolla \\ & & & & E. Townsend  \\ \hline

            Obser. UEPG         & -25 05 22.671 & 406           &  N/A        &   \\ 
            Paraná              & -50 05 56.790 & Raptor Merlin &  N/A        &  M. Emilio \\ 
            Brazil              & 923.0         & Clear         & Overcast  &   \\ \hline

            Sapopema            & -23 51 09.024 & 305           &  N/A        &   \\ 
            Paraná              & -50 39 29.794 & QHY174M-GPS   &  N/A        &  C. Pereira \\ 
            Brazil              & 759.0         & Clear         & Overcast  &  L. A. Pereira \\ \hline

            UNIVAP              & -23 12 30.022 & 300           &  N/A        &  L. Brito, V. Lattari \\ 
            São Paulo           & -45 57 49.407 & Raptor Merlin &  N/A        &  Wandeclayt M. \\ 
            Brazil              & 600.0         & Clear         & Overcast  &  I. Rodrigues \\ \hline

            Gemini South        & -30 14 26.308 & 8,000         &  N/A        &   \\ 
            Cerro Pachón        & -70 44 11.580 & Zorro         &  N/A        &  G. Benedetti-Rossi \\ 
            Chile               & 2,715.4       & Clear         & Overcast  &   \\ \hline

            OACEP               & -25 20 57.494 & 305           &  0.08     &  E. Gradovski \\ 
            Paraná              & -49 21 55.050 & WATEC 910HX   &  N/A        &  J. Antunes \\ 
            Brazil              & 1,055.0       & Clear         & Overcast  &  E. Fonseca, J. Carlos \\ \hline

            OAM - IAG/USP       & -23 00 13.680 & 406           &  N/A        &   \\ 
            São Paulo           & -46 57 52.920 & U9000 APOGEE &  N/A        &  M. Fidêncio Neto \\ 
            Brazil              & 870.0         & Clear         & Overcast  &   
            \\ 
            \hline
            \multicolumn{5}{l}{\textbf{VLT:} Very Large Telescope; \textbf{SPECULOOS-S:} Search for habitable Planets EClipsing ULtra-cOOl}\\
            \multicolumn{5}{l}{Stars - South; \textbf{OUC:} Observatory of the Catholic University; \textbf{CASLEO:} Complejo Astronómico El}\\
            \multicolumn{5}{l}{Leoncito; \textbf{NTT:} New Technology Telescope; \textbf{UEPG:} Universidade Estadual de Ponta Grossa;}\\
            \multicolumn{5}{l}{\textbf{UNIVAP:} Universidade do Vale do Paraíba; \textbf{OACEP:} Observatório Astronômico do Colégio Esta-}\\
            \multicolumn{5}{l}{dual do Paraná; \textbf{OAM - IAG/USP:} Observatório Abrahão de Moraes, Instituto de Astronomia, } \\
            \multicolumn{5}{l}{Geofísica e Ciências Atmosféricas, Universidade de São Paulo.}                    \end{tabular}
            \end{table}

\setcounter{table}{1}
    \begin{table}[ht]
        \footnotesize
        \centering
        \caption{[Cont.] Observation circumstances for all the observatories involved in the September 10, 2023 event.}
        \vspace{4mm}
                \begin{tabular}{c c c c c} \hline \hline
                                                &   \textbf{Latitude ($^\circ$ ' '')}    &  \textbf{ Telescope (mm)}  &   \textbf{Exposure (s)}    &               \\ 
                    \textbf{Site}                        &   \textbf{Longitude ($^\circ$ ' '')}   &   \textbf{Camera}          &  \textbf{ Cycle (s) }      &   \textbf{Observers}   \\ 
                                                &  \textbf{ Altitude (m)}                &   \textbf{Filter }         &   \textbf{Status }         &               \\ 
                    \hline 
            
            CASLEO              & -31 47 13.2   & 600           &  N/A        &  M. Melita \\ 
            San Juan            & -69 18 24.12  & SBig          &  N/A        &  L. Mammana \\ 
            Argentina           & 2,591.0       & Clear         & Overcast    &   \\ 
            \hline
            Mobile              & -31 52 2.1    & 203           &  N/A      &  R. Leiva \\ 
            La Mostaza          & -71 26 25.1   & QHY174M GPS   &  N/A      &  P. Quitral- \\ 
            Chile               & 55.0          & Clear         & Overcast  &  Manosalva \\ 
            \hline
            
            TRAPPIST-S          & -29 15 16.56  & 600           &  0.500    &   \\ 
            La Silla            & -70 44 21.84  & PL3041-BB     &  N/A        &  E. Jehin \\ 
            Chile               & 2,315.0       & Clear         & Overcast  &   \\ 
            \hline
            

            OPD Zeiss           & -22 32 06.664         & 600       &  N/A         & T.F.L.L. Pinheiro    \\ 
            Minas Gerais        & -45 35 0.329         & iKon      &  N/A          & J. P. Cavalcante \\ 
            Brazil              & 1,864.0               & Clear     & Tech. Failure & G. Ramon         \\   
            \hline

            GOA                 & -20 18 02.0   & 304           & N/A         &         \\ 
            Espirito Santo      & -40 19 02.0   & ASI1600MMPro  & N/A         & M. Malacarne   \\ 
            Brazil              & 24.0          &               & Tech. Failure &     \\ 
            \hline

            IFPR Ivaipor\~a     & -24 15 3.874  & 254           & N/A               & G. Margorti        \\ 
            Paraná              & -51 42 47.354 & QHY174M-GPS   & N/A               & O. Margorti   \\ 
            Brazil              & 1,200.0       & Clear         & Tech. Failure     & A. J. Ortiz    \\ 
            \hline
            \multicolumn{5}{l}{\textbf{TRAPPIST-S:} Transiting Planets and Planetesimals Small Telescope - South; \textbf{GOA:} Gaturamo }\\
            \multicolumn{5}{l}{Observatório Astronômico; \textbf{IFPR:} Instituto Federal do Paraná.}\\
            
            \end{tabular}
            \end{table}




\end{document}